A Biomechanical Modeling Study of the Effects of the Orbicularis Oris Muscle and Jaw Posture on Lip Shape

Ian Stavness[a], Mohammad Ali Nazari[b,c], Pascal Perrier[b], Didier Demonlin[b], Yohan Payan[d]

[a] Department of Bioengineering, Stanford University, Stanford, CA, USA

[b] Speech & Cognition Depart, Gipsa-lab, UMR CNRS 5216, Stendhal University & Grenoble INP, Grenoble, France

[c] Department of Mechanical Engineering, Faculty of Engineering, University of Tehran, Iran

[d] TIMC-IMAG Laboratory, CNRS UMR 5525, University Joseph Fourier, La Tronche, France

Correspondence concerning this article should be addressed to Ian Stavness, Department of Bioengineering, Stanford University, Room S221, 318 Campus Drive, Stanford, CA, 94305, USA. E-mail: stavness@gmail.com

Abstract

Purpose: We used a biomechanical model of the orofacial system, to investigate two issues in speech production: (1) the link between lip muscle anatomy and variability in lip gestures; (2) the constraints of coupled lip/jaw biomechanics on jaw posture in labial sounds.

Method: Our methodology is based on computer simulations with a sophisticated biomechanical model coupling the jaw, tongue and face. First, the influence of Orbicularis Oris (OO) muscle geometry is analyzed. We assessed how changes to the OO anatomical implementation, in terms of depth (from epidermis to the skull) and peripheralness (proximity to the lip horn center), affected lip protrusion, lip spreading and lip area. Second, we evaluated the capability of the lip/jaw system to generate protrusion and rounding, or labial closure for different jaw heights.

Results: (1) A highly peripheral and moderately deep OO implementation is most appropriate for achieving protrusion and rounding; (2) a superficial implementation facilitates closure; (3) protrusion and rounding requires a high jaw position; (4) labial closure is achievable for various jaw heights.

Conclusions: Our results provide a quantitative assessment of how physical characteristics of the speech motor plant can determine regularities and variability in speech articulations across humans and over time.

## Introduction

Variations and regularities found in the sound systems of human languages might be due, at least in part, to the intrinsic properties of the orofacial motor system. Variability in vocal tract anatomy across human populations could have initiated differences in articulatory gestures across languages. Likewise, properties shared by all human orofacial motor systems could have been the basis for common articulatory and motor trends observed in a large number of languages. In this context, models of the orofacial motor system can be used to evaluate the influence of variations in physiological and anatomical properties on articulatory speech gestures. Comparing predictions made with models to data collected from speakers of various languages permits a quantitative assessment of the physiological factors that have potentially influenced the emergence of sound system rules and variability in the languages of the world.

Lip gestures are good candidates for investigating potential links between physiological variability in humans and variability in the sound systems of languages because significant differences in facial muscle morphology are known to exist across subjects. These anatomical variations could explain differences in speech specific lip gestures, such as lip protrusion and lip rounding. In a discussion of anthropophonetic variations, Brosnahan (1961) and Catford (1977) quoted the studies of Huber (1931) and stated that the *risorius* muscle is found in about 20% of Australians and Melanesians, 60% of Africans, 75 to 80 % of Europeans and in 80 to 100% of Chinese and Malays. More recently, Pessa et al. (1998) showed that as many as 22 of their 50 cadaver specimens lacked the risorius muscle. Although based on small samples of data, these studies suggest that such genetic anatomical characteristics could be consistent, or even occur with increasing frequency, over successive sections of populations of the African-European-Asian land mass. Pessa et al. (1998) also found that, in 17 of their 50 specimens, the Zygomaticus major presented a bifid structure with two insertions points. The two insertion points of this muscle could cause the dimple in the cheeks that many people have when smiling (Schmidt and Cohn, 2001). These observations regarding the structure of the Zygomaticus major muscle confirm that significant inter-speaker differences exist in facial muscles. Such anatomical differences are likely to determine variations in face shaping and orofacial gestures in facial expression and speech production.

Speech scientists are gradually accumulating data on possible links between anatomical variability across humans and variations in articulatory and acoustical characteristics of human languages. Ladefoged (1984) showed that differences between the vowel systems of Yoruba, a Niger-Congo language spoken in West Africa, and Italian could have an anatomical basis. Ladefoged noted the existence of small differences in formant values between Yoruba and Italian, which otherwise have very similar 7-vowel systems. He noted that these differences are consistent with anatomical differences generally observed between Africans and Europeans. Ladefoged (1984) noted:

> Some of the differences between the two languages are due to the shapes of the lips of Italian as opposed to Yoruba speakers. […] [W]ith the exception of /i/ and to a lesser extent /e/, the second formant is lower for the Italian vowels than for the Yoruba vowels. These differences are precisely those that one would expect if Yoruba speakers, on the whole, used a larger mouth opening than that used by the Italian. […] The possibility of overall differences in mouth opening is certainly compatible with the apparent facial differences between speakers of Yoruba and Italian. (pp. 85-86)

More recently, Storto & Demolin (to appear) found that in Karitiana, a Tupi language spoken in Brazil, none of their 5 subjects showed lip rounding and protrusion while producing the vowel [o]. Measurements were based on video data (Figure 1). The average first and second formant values for this vowel, measured on the 5 speakers, are respectively 459Hz (n = 250) and 1056Hz (n = 250), which corresponds to F1/F2 values for mid-back or high back vowels in the acoustic space. An auditory perceptual test showed that Portuguese and French speakers, who also have vowel [o] in their vowel systems, correctly identified the Karitiana [o] as the corresponding mid-high back vowel [o] in their languages. In addition, electromyographic (EMG) recordings with surface electrodes placed at the rim of the lips showed no activity when Karitiana speakers produced the vowel [o]. In contrast, EMG measurements with the same electrode placement on Portuguese and French speakers, who have lip rounding and protrusion (e.g., Figure 2), showed clear EMG activity for the same vowel. The vowel [o] is the only back vowel of the Karitiana phonetic system, which lacks the high-back vowel [u]. Karitiana is not unique in this regard, as its vowel system is similar to several other Tupi languages (Storto & Demolin 2012). The Yoruba and Karitiana data suggest that small anatomical differences could create variations that influence the shape of sound patterns found in the world's languages. These variations may be one of the factors explaining so-called *phonetic universals*. Other factors include the categorization of these variations and their cultural transmission across many generations.

On the other hand, regularities in phonetic realizations of the world's languages can also originate from physiological factors, including the muscle arrangements and the inter-articulatory interactions in the orofacial region. A fundamental mechanism of speech production is the coupling between the jaw and the tongue and lips, which determines the fine shaping of the vocal tract. This coupling is the basis for reduplicative babbling and its ontogenetic evolution explains variegated babbling (MacNeilage, 1998). Degrees-of-freedom and constraints in this coupling can influence preferences in syllabic patterns in the world's languages. The capability of the upper and lower lips to deform makes it possible for a subject to produce bilabial stops, which require a closed lip-horn, for a range of jaw positions. Hence, the lips can stay in contact while the jaw moves downward, allowing speakers to use anticipatory strategies and co-articulation in speech production movements. For example, transitions from a bilabial stop toward subsequent opened sounds can be produced with a low jaw position without endangering the correct production of the stop. In a recent study, Rochet-Capellan & Schwartz (2007) investigated the coordination between jaw, tongue tip, and lower lip during repetitions of labial-to-coronal (/pata/) and coronal-to-labial (/tapa/) consonant-vowel-consonant-vowel sequences at increasing speaking rate. They found that when speaking rate increased, there was a general trend for the coronal-to-labial sequences to change toward the labial-to-coronal sequences. From articulatory data, they observed that at slow speaking rates both the bilabial and the coronal consonants were produced at the end of upward movements of the jaw, i.e. each consonantal closure was synchronized with the maximal elevation of the jaw. As speaking rate increased, they observed that both consonants were produced during the same upward movement of the jaw. The coronal closure remained synchronized with the maximal jaw elevation, but the labial closure was produced during the upward movement from the vowel to the coronal consonant, i.e. for a lower position of the jaw. This was made possible by the capability of both lips to deform. In many languages, consonantal clusters with first a labial and then a coronal consonant are significantly more frequent than clusters with first a coronal and then a labial consonant. The observation of labial-coronal ordering has been called the Labial-Coronal effect (see MacNeilage and Davis, 2000). The mechanical properties of the lips could have contributed to the emergence of this, and other patterns, in languages.

The goal of our study was to demonstrate the effect of anatomical factors on lip shape using a biomechanical model of the orofacial system. The model integrates the soft-tissues of the face, tongue and lips with the hard-structures of the maxilla and mandible. Considering these hard structures was important for our analysis because the underlying bone structure and mandible position have a significant effect on the configuration of the lips. While previous biomechanical studies of lip protrusion have used generic models (Kim & Gomi 2007, Nazari et. al 2010), we have adapted the morphology of our model to a particular speaker. This decision was motivated by the fact that we have a large set of experimental data for this speaker, which makes possible a quantitative assessment of the simulations. To create our subject-specific model, we coupled the Nazari et al. 2010 face model with the Stavness et al. 2011a jaw/tongue model. Thus, we report here a first-of-its-kind biomechanical model that incorporates coupling and contact effects between the face, the tongue and a muscle-activated jaw.

Using the face model, we investigated two questions regarding variation and regularity in articulation: how variation in lip musculature is associated with variation in lip gestures, and how the properties of lip biomechanics impose constraints on jaw posture. In the first part of the study, we were interested in determining the effect of *orbicularis oris* muscle geometry on simulated lip protrusion and rounding. Our model-based analysis enables a quantitative assessment of how lip shaping is influenced by variations in the anatomical distribution of the marginalis and peripheralis parts of this muscle. Comparing simulations with data, such as recordings of the production of vowel [o] by Karitiana speakers, could provide evidence for links between anatomical and physiological variations and the diversity of the world's languages. In the second part of the study, we were interested in assessing the extent to which a subject can move his/her jaw up and down while keeping a bilabial closure. The model provides quantitative information about motor equivalence strategies likely used to produce bilabial consonants in anticipation of subsequent syllables.

## Methods

We created simulations of lip gestures, using a biomechanical face-jaw-tongue model, for a range of conditions on lip musculature and jaw posture. We compared the simulations using quantitative measurements of the resulting simulated lip shapes.

### Model

In order to assess the effect of variation in orbicularis-oris (OO) morphology and jaw posture on lip protrusion and lip shape we require a model that has consistent anatomy with a specific speaker and that includes the underlying bony structures of the jaw and skull in addition to the soft-tissues of the face and lips. We created our model (see Figure 3) in  the ArtiSynth biomechanical modeling toolkit (www.artisynth.org; see also Lloyd et al., 2012), by registering and integrating two previously reported reference models.

We used a Computed Tomography (CT) dataset of a single speaker as a means to adapt and co-register these disparate reference models. The speaker was chosen because we have an extensive set of experimental data on his speech movements. The reference 3D Finite-Element (FE) face model was originally built in the ANSYS[®] simulation software (Nazari et al., 2010) and consists of 6342 hexahedral elements arranged into three layers: superficial, middle, and deep. The model had been previously adapted to the speaker's CT data using a segmentation of the interior bone surface and the exterior skin surface (Bucki et al., 2010). The reference jaw-tongue-hyoid bone model (Stavness et al., 2011) combined and registered two reference models to the same CT dataset: a 3D rigid-body jaw-hyoid bone model (Hannam et al., 2008) and a 3D FE tongue model (Gerard et al., 2006; Buchaillard et al., 2009).

While both the jaw-tongue-hyoid model and the face model had previously been adapted to the same CT dataset, the inner-surface of the face and lips did not exactly conform to the outer-surfaces of the mandible and maxilla due to inaccuracies in the original data segmentation and model adaptation. The interaction between the lips and underlying bone surfaces is critical during lip movements as the mandible, maxilla, and dentition provide boundary conditions for lip movements. To improve the fit between the two models, we used a contact-based morphing procedure. We also performed some minor manual editing of the face mesh in order to improve the regularity of the finite-elements after adaptation.

We coupled the dynamics of the face, jaw, tongue, and hyoid bone models by defining attachment constraints between the FE nodes of the face and tongue and the rigid bodies of the jaw and hyoid bone. Attachments between the tongue and jaw-hyoid models have been described previously (Stavness et al., 2011). For the face mesh, we attached a number of inner-surface nodes to adjacent locations on jaw and maxilla rigid-bodies, while leaving unattached nodes in the region of the lips and cheeks. We also attached adjacent surfaces of the tongue and face models near the region of the floor of the mouth. The attachment points are illustrated in Figure 3.

Contact between different articulators is another crucial component of speech production and includes deformable to deformable body contact (between the upper and lower lip) as well as deformable to rigid body contact (between the lips and teeth, and between the tongue and palate and teeth). ArtiSynth supports mesh-based collision detection and contact handling using dynamic constraints. Contact detection was enabled between the face and the jaw and maxilla meshes (including the teeth). Sub-regions of an FE face mesh were defined for the upper and lower lips and used for contact between the lips.

In the reference face model, muscle forces were applied along serial line segments representing the muscle's principle line of action (called *cable elements*, see Nazari et al., 2010). In the current model, muscle mechanics is incorporated with a transverse-isotropic FE material, whereby stress is increased in the fiber direction with muscle activation (Weiss et al., 1996). The OO muscle was defined as a continuous loop of Finite Elements around the lips, as shown in Figure 4. To vary the OO morphology, the size and location of this loop of elements was varied in different simulations, as described in the *Simulations* section. The fiber direction in a finite-element associated with a particular muscle represents the muscle's principle line-of-action and additional stress is applied in that direction within the element during muscle activation. The fiber directions for each individual finite-element associated with the OO muscle were interpolated from a canonical serial line segment representing the OO muscle fibers, as shown by the cyan lines in Figure 5. If an element lies within the region of the cyan loop then its fiber direction is interpolated from the nearby cyan line segments. The interpolation is performed as a weighted average of line segment directions, where the weighting is inversely proportional to the distance between the element and line segments. Whereas, if an element lies outside of the region of the cyan loop, then its direction is set to the direction of the closest line segment of the cyan loop.

The face model was implemented into ArtiSynth using a large deformation FE framework, hexahedral elements with a density of 1040 kg/m$^3$, and a fifth order incompressible

Mooney-Rivlin material with parameters consistent with the Nazari et al. (2010) reference face model ($c_{10}$=2500 Pa, $c_{20}$ = 1175 Pa, and $c_{01}$ = $c_{11}$ = $c_{02}$ = 0 Pa). Incompressibility was implemented using a constraint based mixed u-P formulation (see Stavness et al., 2011 for further details). Rayleigh damping coefficients of $\alpha$ = 19 s$^{-1}$ and $\beta$ = 0.055 s were used for consistency with the reference face model (Nazari et al., 2010). The transverse-isotropic muscle material property was superimposed on the passive isotropic material property due to the uncoupled strain energy formulation proposed by Weiss et al. (1996). Passive stress in the fiber direction increases exponentially with increasing fiber stretch (see Weiss et al., equation 7.2, p. 123). Parameters for the passive fiber behavior were chosen based on parameters for muscle provided by Blemker et al. (2005): $C_3$=0.05, $C_4$=6.6, $C_5$=0, $\lambda^*$=1.4. The maximum active fiber stress was 100 kPa.

**Simulations**

Our primary aim was to determine the effect of OO morphology on simulated lip protrusion and rounding. We were also interested in the effect of jaw posture on lip rounding and protrusion, which we are able to analyze with our first-of-its-kind coupled face-jaw-tongue biomechanical model.

Simulations were performed in ArtiSynth, which allows for fast forward dynamics simulation with dynamic coupling between rigid-body and FE models as well as collision handling. Coupling and collision handling is important for modeling the interactions between the lips and the underlying bony structures. ArtiSynth also provides graphical user interface tools that were used to help with model registration and FE mesh editing.

We achieved simulation times that were much faster than the reference face model in ANSYS. For the full model, with dynamic coupling and contact, each 500 ms simulation required approximately 225 s of simulation time on a 2.2 Ghz Intel Core i7 processor. It has been shown that ArtiSynth can be orders of magnitude faster than ANSYS for similar simulation (Stavness et al. 2011).

Each simulation was 500 ms in duration: muscle activation for the OO muscle increased linearly over duration of 400 ms and held the final activation for 100 ms. In all simulations, muscle activation was increased uniformly from 0 to 50 % of the maximum possible activation, which corresponds to an active muscle stress of 50 kPa. This level of final activation was chosen to ensure numerical convergence in all simulations while generating lip displacements of realistic amplitudes. Each simulation reached an equilibrium position by 500 ms.

**Deepness and Peripheralness.** The OO muscle was modeled as a continuous loop of finite-elements. In order to assess the effect of deepness, we ran simulations with the OO muscle located only in the deep (D), or in the middle (M), or in the superficial (S) layer of the face mesh, where deep is closer to the skull and superficial is closer to the skin surface (see Figure 3). In order to assess the effect of peripheralness, we varied the radius of the OO muscle loop (centered in the middle of the lip horn) from smaller radius (more medial) to larger radius (more peripheral) in four sizes: 1, 2, 3, and 4 (as shown in Figure 4).

**Upper versus Lower OO Peripheralness.** To investigate the variation of OO size in more detail, we also performed simulations of different peripheralness for the upper versus the lower portion of the OO muscle. The structure/geometry of the FE face mesh is such that the lower portion of the OO muscle is more peripheral to the lower lip than the upper portion of the OO muscle is to the upper lip (by inspection of Figure 4). Therefore, we tested different relative peripheralness of the upper versus lower parts the OO muscle. In these simulations, the OO muscle was located in all tissue layers: deep, middle, and superficial.

**Lip Rounding with Jaw Lowering.**  To investigate the effect of jaw lowering on lip rounding and protrusion, simulations were performed with synergistic activation of OO and jaw lowering muscles (anterior-belly of the digastric (ABD) and lateral pterygoid (LP) muscles). The jaw model was dynamically coupled to the FE models of the tongue and face, and therefore we were able to simulate the biomechanical effect of the coupled system. We chose the OO muscle configuration from the above simulations that best matched the speaker's lip protrusion (shown in Figure 2).

**Lip Closure with Jaw Lowering.**  In order to investigate the extent to which lip closure is compatible in the model with jaw lowering, different configurations of OO muscle were evaluated. Based on the results of our *Deepness vs. Peripheralness* simulations, we expected that lip closure could be achieved with the superficial portions of the OO muscle. We varied the different amount of peripheralness in OO needed to achieve closure with two different degrees of jaw lowering.

## Lip Shape Metrics

In order to quantify the effect of OO morphology on lip shape, we chose lip measurements similar to those proposed by Abry and Boë (1986) and used in Nazari et al. (2011, see Figure 7 and 8, pg. 151). Forward lip protrusion was characterized by the anterior displacement of the most anterior point on the upper and lower lips, as illustrated in Figure 6 (left panel). We calculated the displacement as the difference in position between the most anterior flesh-point in the protruded posture from the most anterior flesh-point in the rest posture (for both the upper and lower lip). The anterior points in each posture may be different flesh points as the lips may rotate during protrusion. Positive displacement corresponds to anterior protrusion relative to rest posture. The lip opening was characterized by the shape of the opening in an orthographic projection of the frontal view of the model. The opening space was segmented from the frontal projection image and its width, height, and area are measured as shown in Figure 6 (right panel).

## Results

### Deepness and Peripheralness

Simulation results of lip protrusion for different configurations of OO muscle geometry are plotted in Figure 7 for the same level of activation of the active muscle elements. Quantitative lip measurements for the simulations are reported in Table 1. The results show that more peripheral OO implementations are associated with larger protrusion, independent of deepness, with one exception in the superficial layer (see below). However, the degree of deepness influences the covariation of protrusion and lip area. For a deep OO implementation, peripheralness and protrusion are systematically associated with larger lip width and lip height, and therefore with larger lip area. For a superficial implementation, peripheralness is also associated with larger lip area, mainly due to an increase in lip width. For a middle OO implementation, the influence of peripheralness is different: we observe a non-linear variation in lip height, lip width, and lip area with peripheralness. From peripheralness 1 to peripheralness 3, lip area increases, mainly because of the increase in lip height; but lip area decreases from peripheralness 3 to peripheralness 4 due to the combined decrease in lip height and lip width.

The prototypical characteristic of the protrusion/rounding gesture, as observed in rounded vowels such as /u/ or /o/, is a significant amount of lip protrusion associated with a small lip opening area. In the conditions of our simulations, the protrusion/rounding gesture is most effectively produced for the most peripheral implementation of the OO muscle in the middle layer of the face tissues (especially for the lower lip). Interestingly, an implementation in the superficial layer cancels the influence of peripheralness on protrusion for the upper lip (see Table 1*Pu*). Hence, a superficial implementation of the OO seems to be particularly inappropriate to the generation of protrusion and rounding.

The results show also that for a given degree of peripheralness, a superficial location enables efficient closure, and a deep location reduces the impact of OO activation on closure. Hence in the absence of protrusion requirement, a superficial implementation facilitates lip closing gestures.

Except in one case (middle and most marginal implementation of the OO), the lower lip shows larger protrusion than the upper lip. This may be a subject-specific property due either to the subject lip volume (i.e. a lower lip volume larger than the upper lip volume) or to the fact that the lower OO muscle is more peripheral than the upper OO muscle in the finite-element mesh, as shown in Figure 4. This issue is investigated below.

**Upper versus Lower OO Peripheralness**

The results of the previous section show differences in protrusion amplitude between the upper and the lower lips. This might be due to differences in the peripheralness of the OO implementation. Simulation results for lip shapes obtained with differential upper versus lower OO peripheralness are plotted in Figure 8 and quantitative measures are reported in Table 2. We assumed an implementation of the OO in all layers (superficial, middle and deep) together. We chose to test more peripheral implementations for the upper portion of OO (ring 3 and 4, as compared to ring 2 and 3 for the lower portion) because the structure/geometry of the FE face mesh is such that the lower portion of the OO muscle is more peripheral to the lower lip than the upper portion of the OO muscle is to the upper lip (by inspection of Figure 4).

In general, the protrusion amplitudes are much larger (between 1 and 3mm) for both parts of the lips than in the previous section. This is due to the fact that here all layers of the OO were activated together, while in the *Deepness and Peripheralness* simulations each layer was activated separately. Apart from this side-effect, the results demonstrate that differential protrusion of the upper versus lower lip is determined by the peripheralness of the upper versus lower OO muscle fibers. Consistent with the results of the previous section, the protrusion of the lower lip is systematically larger than the protrusion of the upper lip.

As expected from the results of the previous section, a more peripheral implementation of the OO muscle in one part of the lips increases the protrusion of the same part. However, a more peripheral implementation of the upper part reduces the protrusion of the lower part. A consequence of this phenomenon is that the smallest difference in protrusion between the upper and lower lips is obtained for a peripheral implementation of the OO in the upper lip (4th radius) and a marginal implementation in the lower lip (2nd radius), while the largest difference is obtained for a marginal implementation in the upper lip (3rd radius) and peripheral implementation in the lower lip (3rd radius).

Ideally, lip rounding is associated with a small width and a reasonably small area (between 20 and 30mm$^2$). Smallest widths are obtained for the more marginal implementation of the OO in the lower lip (2nd radius). For this lower radius, the more appropriate lip area is obtained for the less peripheral implementation in the upper lip (3rd radius). This configuration (2nd lower radius and 3rd upper radius) provided the best tradeoff between lip rounding and protrusion. We also found that this configuration provided the best qualitative match to the subject's data (Figure 2). This best case OO configuration was chosen for simulations below on lip protrusion during jaw lowering.

**Lip Rounding with Jaw Lowering**

We evaluated the degree to which lip rounding is compatible with jaw lowering using simulations with increasing degrees of jaw lowering during OO activation. Results are plotted in Figure 9 and quantitative lip measurements reported in Table 3. The best OO configuration obtained in the previous section was used: $2^{nd}$ lower radius and $3^{rd}$ upper radius with an implementation of the OO in the three layers together (superficial, middle and deep). Jaw lowering distance was measured as the distance between the lower and upper mid-incisor points. Increase in jaw lowering muscle activation lowers the mandible, and, obviously, it causes increased lip opening. The backward movement of the anterior part of the jaw, associated with jaw lowering, moves the lower lip backwards. The upper lip position also varies with jaw position, but is affected less than the lower lip. These simulation results show that the rounding/protrusion gesture is rather sensitive to variation in jaw height and suggest having a high jaw position is a requirement for the achievement of a correct protrusion and rounding lip gesture.

**Lip Closure with Jaw Lowering**

We evaluated the degree to which lip closure is compatible with jaw lowering. The results from the *Deepness and Peripheralness* simulations suggested that lip closure could be achieved with the superficial portions of the OO muscle. Indeed, we could achieve lip closure, similar to lip shapes in bilabial consonants /b/ or /p/, for a low jaw posture by activating both the peripheral and marginal portions of the OO muscle in the superficial layer (S1+S2+S3+S4). The results are plotted in Figure 10 and quantitative lip measurements reported in Table 4. The additional recruitment of middle, marginal portion (M1) achieves lip closure with a very low jaw posture. The peripheral OO activation provided the required closure of the lips by downward movement of the upper lip and upward movement of the lower lip. Notably, we also observe coupling effects between the face and jaw: activation of OO to achieve lip closure induces slight jaw closure. These simulations demonstrate that, contrary to lip rounding, lips closure is compatible with variable jaw heights.

## Discussion

The properties and structure of sounds in human languages, including diachronic evolution, arrangement into sequences, (co)articulation, and variability of acoustic and articulatory correlates, are the results of a complex combination of influences. These influences arise from various factors, including the intrinsic physical properties of the speech production system, basic motor control principles of human skilled movements, the intrinsic properties of the auditory and visual perception systems, memorization capabilities in humans, social factors, environmental factors, and communication efficiency principles. A major limitation of experimental studies that aim to explain how sound systems in the world's languages are structured, varied, and evolved, is the difficulty of disentangling these different influences in the signals recorded from human speakers.

In this context, the use of computational models to represent the various processes that contribute to language structure is a potentially fruitful approach. Models are designed to be a simplified representation of reality. In addition, modeling isolated subsystems of a complex process, without accounting for their interactions, provides only a partial view into the whole process. However, these models can give a clear picture of the constraints that each subsystem exerts on the whole process. The goals of the present study were to assess the constraints that orofacial biomechanics exerts on speech production gestures and to provide an interpretation of these constraints in the context of regularity and variability of the sound systems in the world's languages. We focused specifically on the lip gestures as a prototypical case. Toward this aim, we used a sophisticated realistic 3D biomechanical model of the whole orofacial motor plant including the jaw, the tongue and the face (Hannam et al., 2008; Buchaillard et al., 2009; Nazari et al., 2010; Stavness et al., 2011).

Given that variability in orofacial muscles anatomy has been found across humans, we tested the influence of plausible variability in the anatomical implementation of the orbicularis-oris (OO) muscle on lip shapes. This muscle plays a central role in the production of the lip protrusion and rounding gestures (Nazari et al., 2011). These gestures are essential for rounded vowels such as /u/, /o/ or /y/, and for closing labial gestures, which represent the key articulatory feature of bilabial stops such as /b/ and /p/. Our approach consisted of measuring the variability of key parameters of the lip shape gesture when the OO anatomical implementation was systematically varied. A potentially confounding factor is that motor control strategies can be adapted across speakers, or for the same speaker across conditions, in order to achieve the same motor goal with different configurations of the motor plant (see for example Hughes & Abbs, 1976). However, we intentionally did not consider this possibility in our study and used the same level of muscle activation for all the tested anatomical OO implementations. This limited scope allowed us to evaluate the intrinsic influence of anatomy, independent of any possible adaptation of motor control strategies.

Variability was tested in terms of depth of the OO anatomical implementation, from superficial (i.e. close to the skin) to deep (i.e. close to the maxillary bones), and in terms of distance from the center of the lip horn, from marginal (i.e. close to the center) to peripheral.

Differences were also considered in the implementation between the upper and the lower lips. Not surprisingly, it was found that anatomical variability has a noticeable impact on lip shaping. When considering similar implementations in the upper and the lower lips, we observed general trends such that a superficial location facilitates closure, a deep location acts against closure, and a peripheral location causes protrusion and aperture. However, some non-monotonous relations were also observed for the most peripheral OO implementations. Interactions were also found between deepness and peripheralness, such that a largely peripheral implementation with an intermediate deepness is the most appropriate implementation for the efficient achievement of the protrusion and rounding gesture.

The evaluation of some differences in the OO implementation between the upper and the lower lip revealed that a very peripheral implementation in the upper lip associated with an intermediate implementation of the lower lip generates the best protrusion and rounding gesture. Conversely, a superficial implementation is not well adapted for the production of this gesture, because it does not facilitate protrusion.

In the context of the study of sound systems in the world's languages these results can be interpreted as follows. If anatomical differences in the orbicularis-oris anatomy exist across groups of humans, which are consistent with the hypotheses underlying our work, it can be expected that more rounded and protruded sounds exist in groups of humans where the implementation is more peripheral and reasonably deep in both lips. Obviously, since languages are under the influence of numerous factors, we do not assert that all the languages in these groups of humans would have these characteristics. However, these findings could explain a general trend in these languages as concerns protruded and rounded vowels. Considering in turn the example of Karitiana speakers, our modeling results suggest that a potential explanation for the limited amount of protrusion and the relatively large aperture of the lip horn during /o/ could lie in an OO implementation that is more marginal and/or deep. Future EMG studies are planned to clarify the location of the muscle fibers that are activated during the production of vowel /o/ in Karitiana.

The second part of our study aimed to evaluate the extent to which achieving correct lip gestures, including protrusion/rounding and labial closure, are compatible with variations in jaw height. In the majority of the languages, protruded and rounded labial vowels, as well as bilabial stops, are associated with high jaw positions. We aimed to assess whether a high jaw position is a strict biomechanical requirement for theses sounds or whether there exists freedom to lower the jaw without endangering production of the correct lip gesture. Our simulations suggest that jaw height is indeed a strong requirement for the achievement of a correct protrusion and rounding gesture. This could explain the fact that the majority of rounded vowels are high vowels. However, our simulations tend to show that a high jaw position is not a requirement for bilabial closure. In our model, a reasonable activation of the superficial layer of the OO in its marginal and peripheral parts enables lip closure even if the distance between the upper and the lower incisors is as large as 1cm. Such a large freedom in jaw positioning can certainly be used to plan sequences of speech gestures in order to find the most appropriate jaw trajectories to achieve of sequence of articulatory goals within a short time interval, or to achieve anticipated articulatory goals with co-articulation. The simulated freedom in jaw posture during bilabial stops is consistent with the observed preference for the Labial-Coronal order in sequences of consonants in the world's languages. Rochet-Capellan and Schwartz (2007) (Figure 5B) suggested that this effect arises when the bilabial stop is produced for a low jaw position during the jaw upward movement from the preceding vowel to the subsequent coronal stop. Our simulations confirm that this is indeed possible, even for relatively low jaw positions and plausible magnitude of muscle activations.

## Conclusions

Our study demonstrates the utility of a realistic model of orofacial biomechanics in the analysis of variation and regularity in articulatory patterns. We found evidence for potential links between variability in the anatomy of the lips and variability in articulatory and acoustical characteristics of speech sounds. We also associated the biomechanical constraints of the lips with regularities in articulatory gestures. We demonstrated that the deformation capabilities and muscle activations of the lips allow for a certain amount of freedom in jaw positioning during bilabial stops. This freedom could be used for co-articulation planning in bilabial stops-vowel-coronal stops-vowel sequences at fast speaking rate. Conversely, we also demonstrated that protrusion and rounding of the lips requires high jaw positions.

These results contribute to a better understanding of the influences under which the languages of the world structured themselves and varied diachronically. We have shown that physical differences and regularities of the speech motor plant are factors that may influence the emergence and the evolution of speech sounds.

## Acknowledgements

The authors thank the ArtiSynth team at the University of British Columbia for making the simulation software available, Pierre Badin at Gipsa-lab for providing the CT data used to adapt the model as well as for Figure 2.

Table 1

*Quantitative measurements for Deepness versus Peripheralness simulations*

| *Pu*: Upper Lip Protrusion (mm) | | | | |
| --- | --- | --- | --- | --- |
| | Peripheral Radius | | | |
| Depth | 1 | 2 | 3 | 4 |
| Deep | 0.6 | 1.9 | 2.9 | 3.9 |
| Middle | 0.3 | 1.4 | 2.0 | 2.7 |
| Superficial | 0.3 | 0.0 | 0.1 | 0.2 |

| *Pl*: Lower Lip Protrusion (mm) | | | | |
| --- | --- | --- | --- | --- |
| | Peripheral Radius | | | |
| Depth | 1 | 2 | 3 | 4 |
| Deep | 1.6 | 3.3 | 4.4 | 4.8 |
| Middle | 0.0 | 2.5 | 4.5 | 4.8 |
| Superficial | 0.6 | 0.9 | 1.5 | 1.5 |

| *W*: Lip Opening Width (mm) | | | | |
| --- | --- | --- | --- | --- |
| | Peripheral Radius | | | |
| Depth | 1 | 2 | 3 | 4 |
| Deep | 14.8 | 21.9 | 25.7 | 27.1 |
| Middle | 10.9 | 9.6 | 14.4 | 12.9 |
| Superficial | 0.0 | 0.0 | 8.4 | 9.3 |

| *H*: Lip Opening Height (mm) | | | | |
| --- | --- | --- | --- | --- |
| | Peripheral Radius | | | |
| Depth | 1 | 2 | 3 | 4 |
| Deep | 2.3 | 3.0 | 4.1 | 4.9 |
| Middle | 1.0 | 1.6 | 3.2 | 2.6 |
| Superficial | 0.0 | 0.0 | 0.8 | 0.7 |

| *A*: Lip Opening Area (mm$^2$) | | | | |
| --- | --- | --- | --- | --- |
| | Peripheral Radius | | | |
| Depth | 1 | 2 | 3 | 4 |
| Deep | 22.0 | 33.1 | 59.5 | 74.6 |
| Middle | 6.3 | 7.7 | 26.5 | 18.0 |
| Superficial | 0.0 | 0.0 | 3.3 | 3.6 |

Table 2

*Quantitative measurements for Upper versus Lower Lip Peripheralness simulations*

| *Pu*: Upper Lip Protrusion (mm) | | |
|---|---|---|
| | Upper OO Peripheral Radius | |
| Lower OO Peripheral Radius | 3 | 4 |
| 2 | 3.8 | 4.3 |
| 3 | 4.3 | 5.4 |

| *Pl*: Lower Lip Protrusion (mm) | | |
|---|---|---|
| | Upper OO Peripheral Radius | |
| Lower OO Peripheral Radius | 3 | 4 |
| 2 | 6.6 | 6.1 |
| 3 | 7.8 | 7.6 |

| *W*: Lip Opening Width (mm) | | |
|---|---|---|
| | Upper OO Peripheral Radius | |
| Lower OO Peripheral Radius | 3 | 4 |
| 2 | 13.8 | 11.6 |
| 3 | 16.4 | 16.4 |

| *H*: Lip Opening Height (mm) | | |
|---|---|---|
| | Upper OO Peripheral Radius | |
| Lower OO Peripheral Radius | 3 | 4 |
| 2 | 3.4 | 2.3 |
| 3 | 5.1 | 3.9 |

| *A*: Lip Opening Area ($mm^2$) | | |
|---|---|---|
| | Upper OO Peripheral Radius | |
| Lower OO Peripheral Radius | 3 | 4 |
| 2 | 25.3 | 11.9 |
| 3 | 51.4 | 31.4 |

Table 3

*Quantitative measurements for OO activation with varying degrees of jaw lowering*

| Jaw Lowering (mm) | Upper Lip Protrusion (mm) | Lower Lip Protrusion (mm) | Lip Opening Width (mm) | Lip Opening Height (mm) | Lip Opening Area (mm$^2$) |
|---|---|---|---|---|---|
| 10 | 4.3 | 8.3 | 14.4 | 3.0 | 19.1 |
| 16 | 3.0 | 4.8 | 18.5 | 5.6 | 63.4 |
| 21 | 3.7 | 4.1 | 22.4 | 8.6 | 122.2 |

Table 4

*Quantitative measurements for 10% and 20% jaw lowering muscle activation with different OO*

*configurations*

| 10 % jaw lowering muscles activation (~10mm jaw lowering) | | | | |
|---|---|---|---|---|
| OO Muscles | Upper Lip Protrusion (mm) | Lower Lip Protrusion (mm) | Lip Opening Width (mm) | Lip Opening Height (mm) | Lip Opening Area (mm$^2$) |
| Rest | 0.0 | -1.6 | 37.2 | 7.6 | 175.6 |
| S1 | 0.3 | -0.9 | 33.7 | 5.5 | 95.2 |
| S12 | 0.3 | -0.1 | 21.8 | 3.6 | 32.4 |
| S123* | 0.3 | 1.0 | 0.0 | 0.0 | 0.0 |
| S1234* | 0.5 | 1.9 | 0.0 | 0.0 | 0.0 |
| S1234+M1* | 1.0 | 2.6 | 0.0 | 0.0 | 0.0 |

| 20 % jaw lowering muscles activation (~16mm jaw lowering) | | | | |
|---|---|---|---|---|
| OO Muscles | Upper Lip Protrusion (mm) | Lower Lip Protrusion (mm) | Lip Opening Width (mm) | Lip Opening Height (mm) | Lip Opening Area (mm$^2$) |
| Rest | 0.0 | -2.0 | 37.9 | 10.3 | 258.9 |
| S1 | 0.3 | -1.3 | 35.9 | 7.8 | 166.8 |
| S12 | 0.3 | -0.5 | 33.7 | 6.0 | 88.5 |
| S123 | 0.3 | 0.6 | 20.3 | 3.6 | 27.4 |
| S1234 | 0.2 | 1.3 | 9.0 | 1.3 | 3.4 |
| S1234+M1* | 0.6 | 2.1 | 0.0 | 0.0 | 0.0 |

*Note.* Cases where lip closure is achieved are denoted by *.

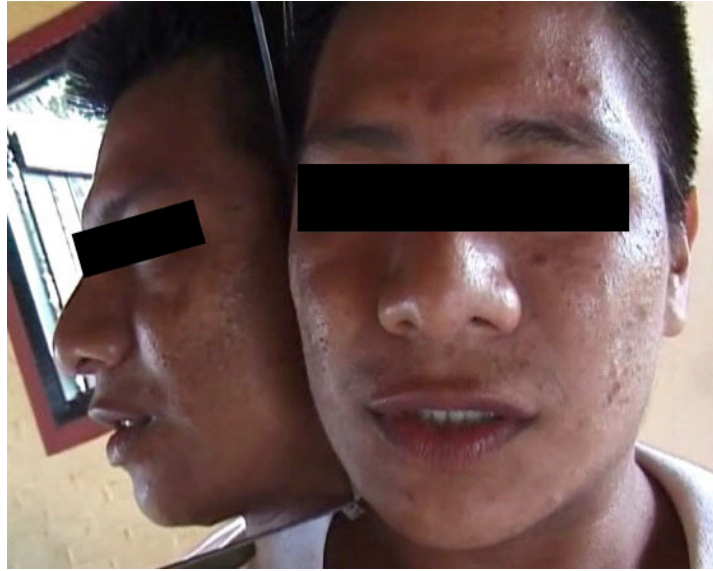

*Figure 1.* Lateral and frontal views of the face of a Kartiana speaker producing the vowel

[o] in the word [koçotl] 'sweet'. The frame was chosen from the corresponding acoustic

recording. The position corresponds to the middle part of the first vowel in the word. It can be observed that the lip shape does not match the classical patterns of protruded and rounded lips; cf. Figure 2.

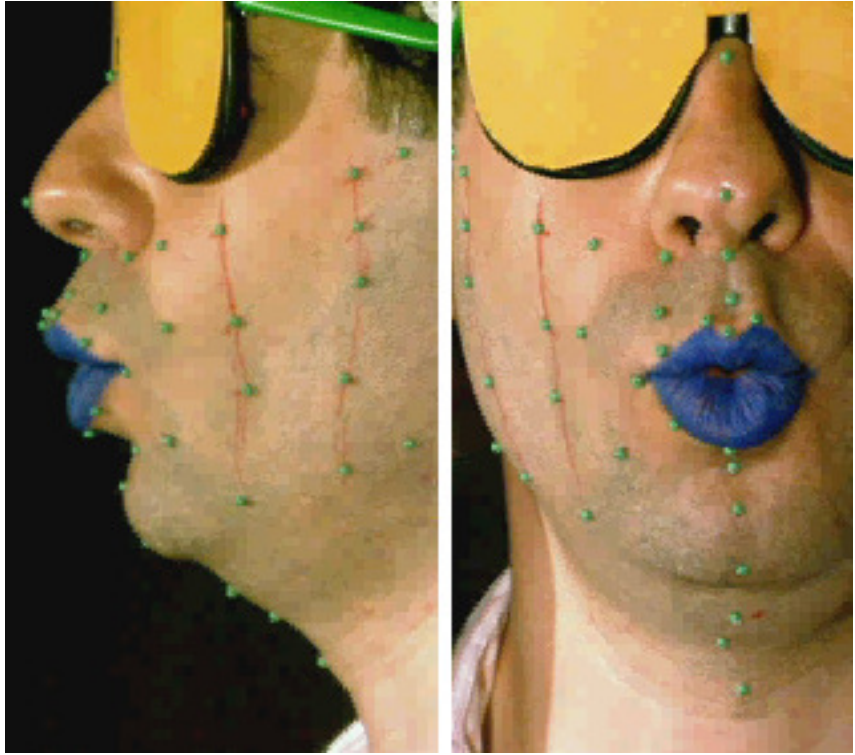

*Figure 2.* Lateral (left) and frontal (right) views of the face of a French speaker

producing the French vowel [u]. The frame was chosen from the corresponding acoustic

recording. A rounded and protruded lip shape can be observed. (Courtesy of Pierre

Badin, Badin et al., 2002.)

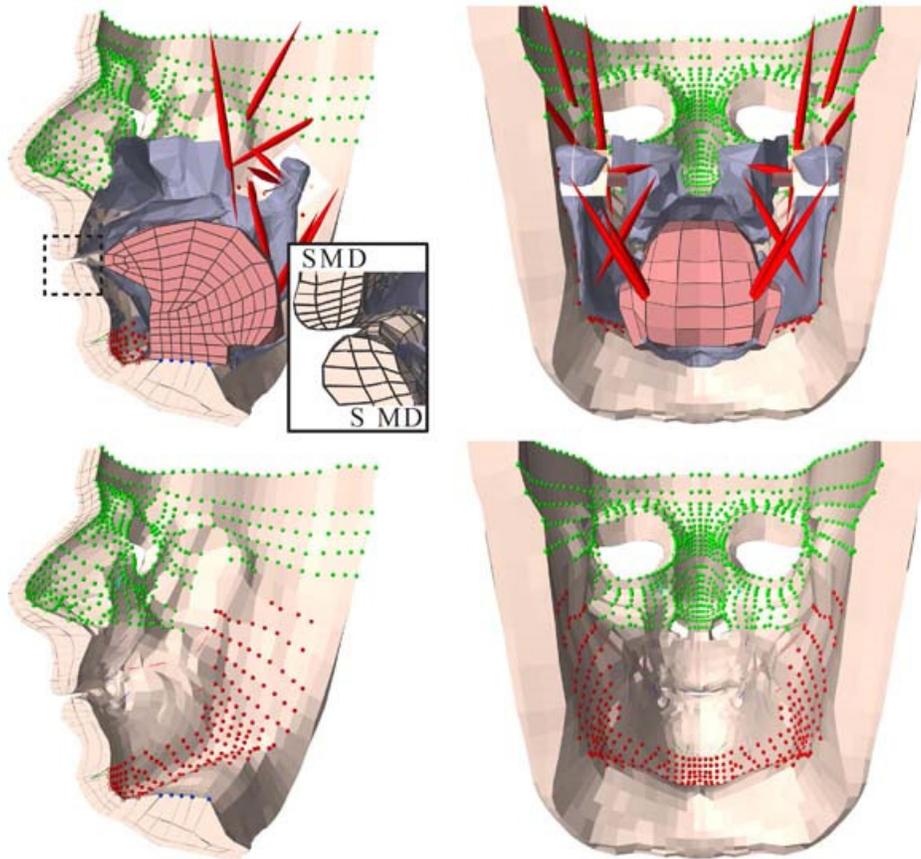

*Figure 3.* Upper panel: Sagittal cut-away (left) and posterior (right) views of the dynamic model that integrates the face, jaw, tongue, and hyoid bone.  A close-up view of the lips (inset, left) shows the superficial (S), middle (M), and deep (D) layers of the three-layer FE mesh. Bottom panel: Attachment points are shown between the face-skull (green points), face-jaw (red points), and face-tongue (blue points).

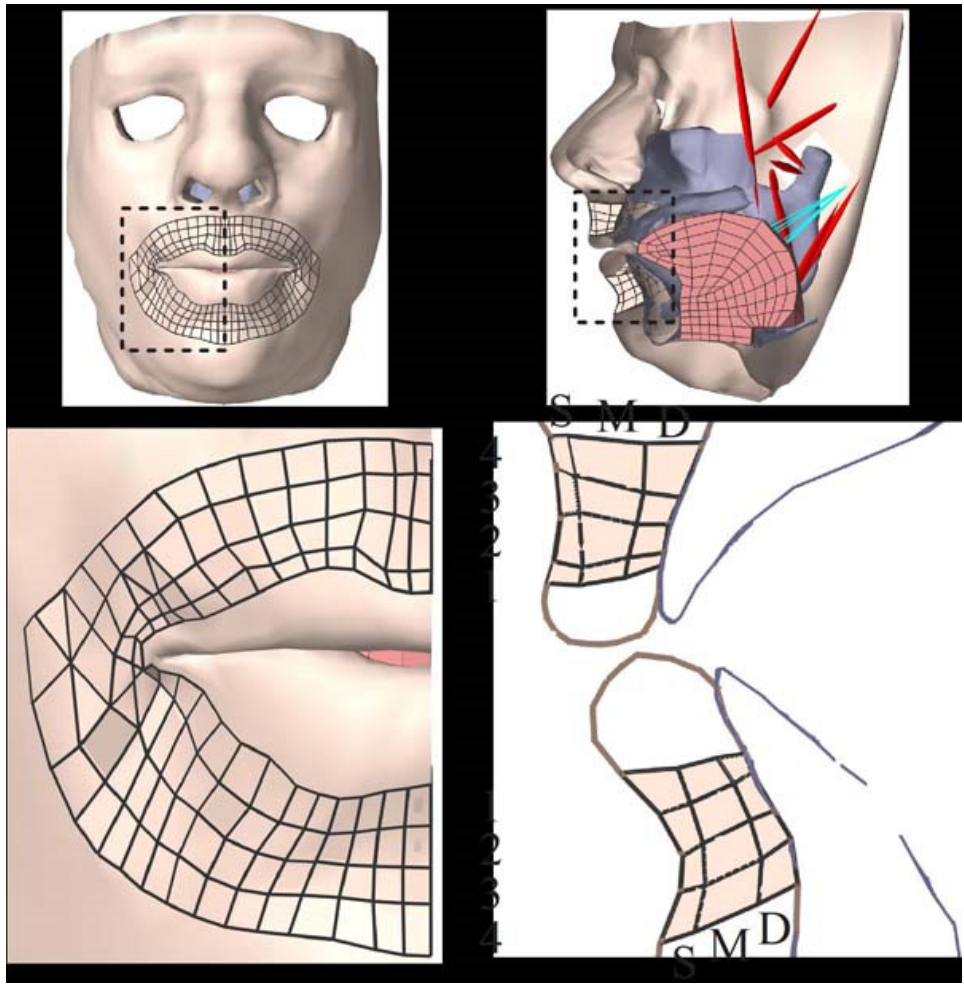

*Figure 4.* Frontal (left) and lateral (right) views of the face model showing the OO muscle elements organized into different peripheral loops from marginal to peripheral (1, 2, 3, 4), and into different depth layers in the mesh, superficial (S), middle (M), and deep (D).

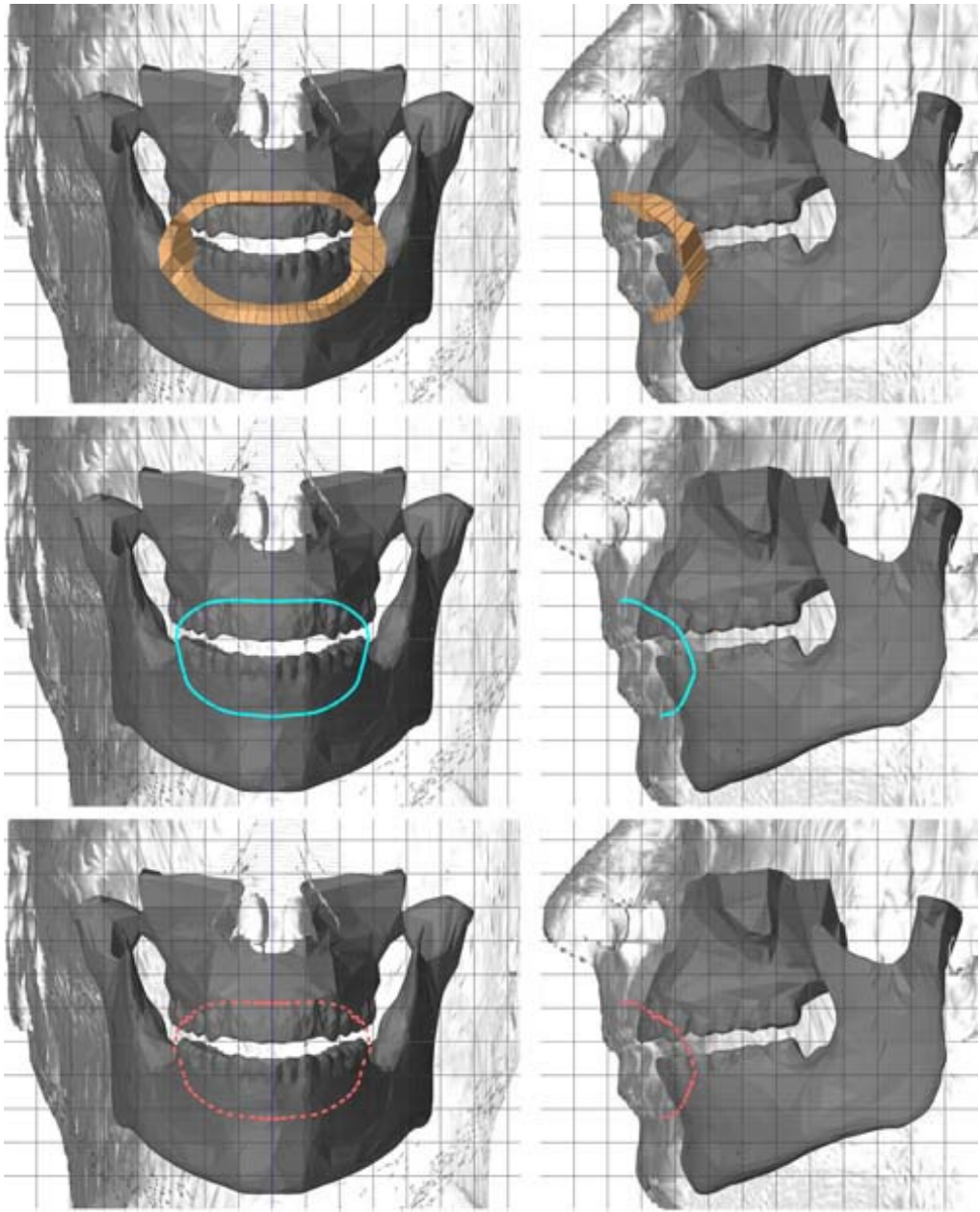

*Figure 5.* Frontal and lateral views of the face model showing the OO muscle elements for M1 radius, M layer (left), and OO muscle element directions (right). Muscle element directions (red lines) are interpolated from the putative OO fiber direction (cyan lines). Grid spacing is 10mm.

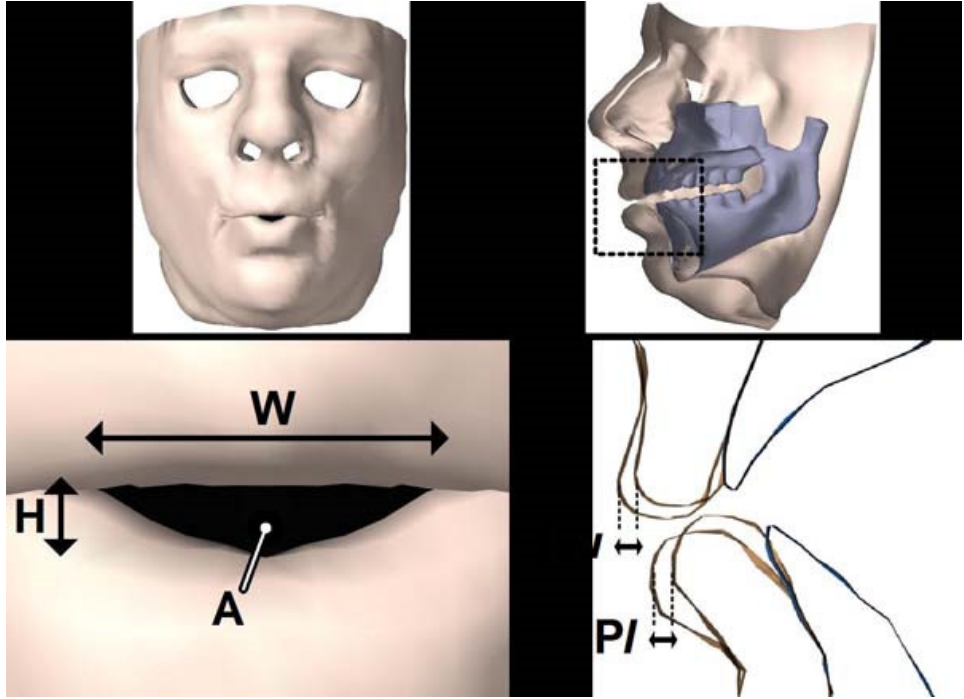

*Figure 6.* Lip protrusion and shaping metrics (inspired by Abry & Boë, 1986). Frontal view (left): lip opening  width (**W**), height (**H**), and area (**A**). Lateral view (right): upper/lower lip anterior protrusion (**Pu/Pl**).

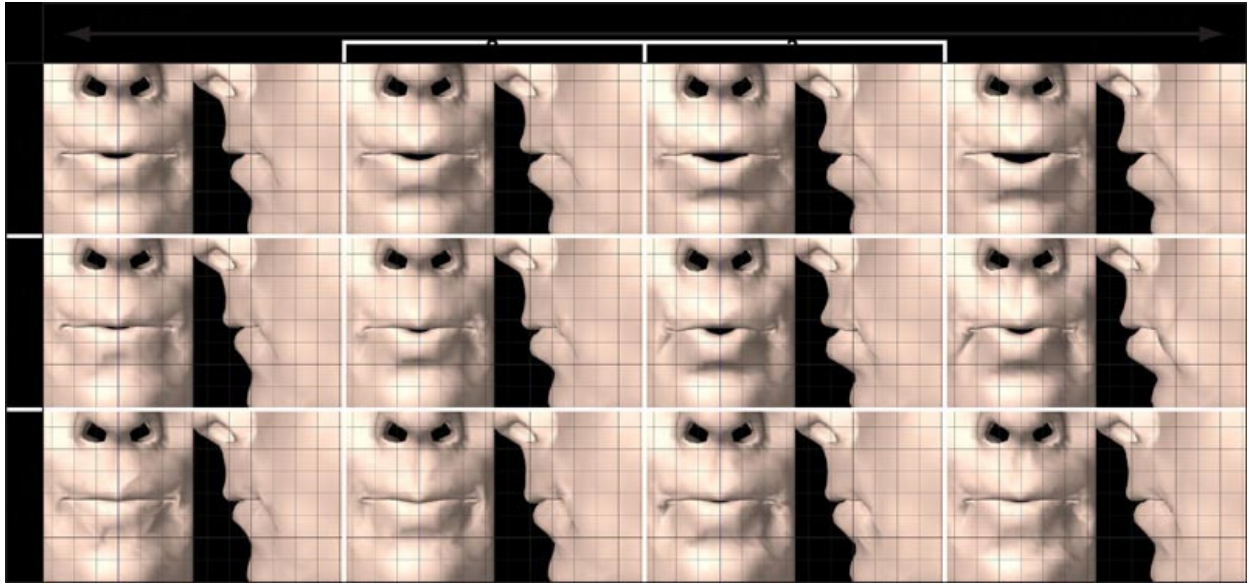

*Figure 7.* Simulation results for different OO muscle deepness and peripheralness. Superficial placement of OO resulted in more closure, whereas deep placement resulted in more opening. Peripheral placement of OO resulted in general in more protrusion and more aperture than marginal placement.

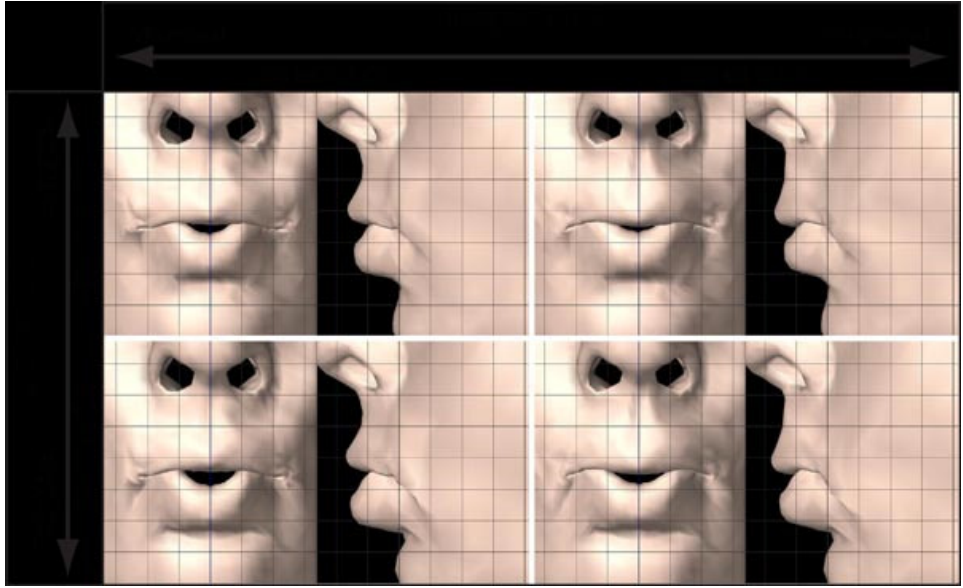

*Figure 8.* Simulation results for different peripheralness of the upper versus the lower

portion of the OO muscle.

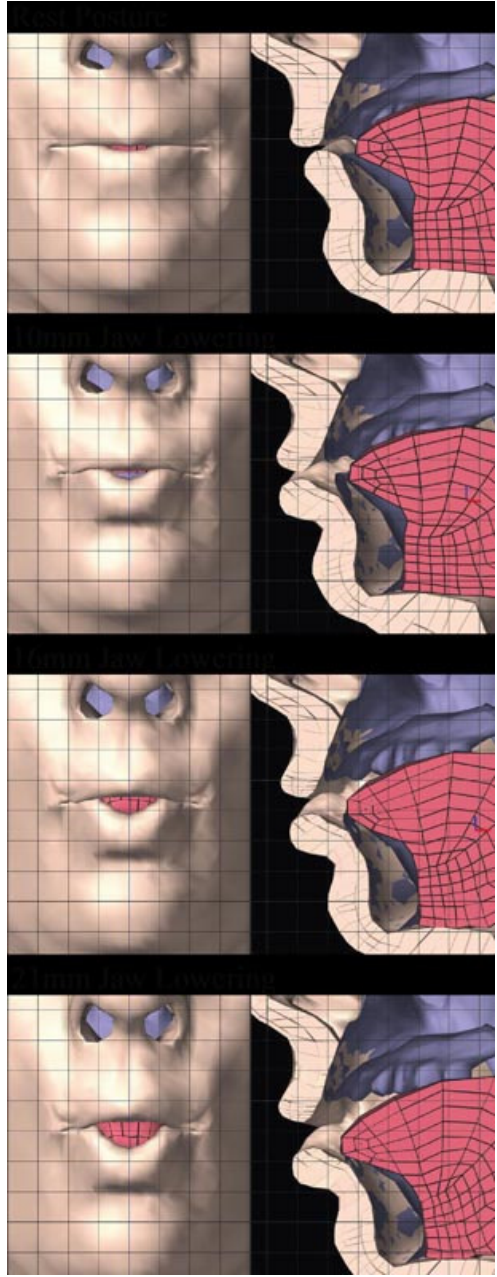

*Figure 9.* Simulated lip rounding and protrusion for different levels of jaw lowering. Jaw lowering increases lip opening and reduces lip protrusion.

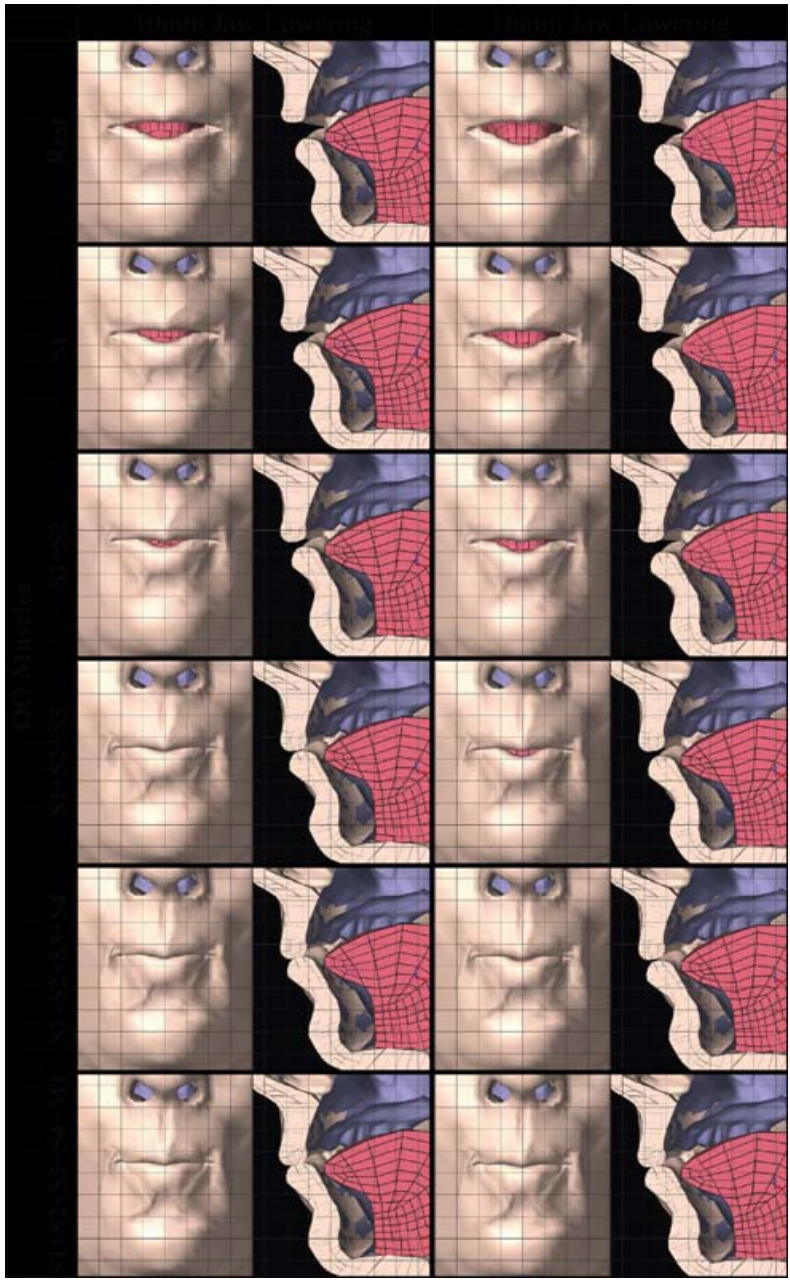

*Figure 10.* Simulated lip closure with different OO configurations for two different levels of jaw lowering. 10% activation of jaw openers resulted in approximately 10mm jaw lowering (left panels), while 20% activation achieved approximately 16mm lowering (right panels). Lip closure was achieved by activating the superficial layer of the OO muscle.